# Incorporation of Si atoms into CrCoNiFe high-entropy alloy: a DFT study

S. Assa Aravindh[1], Andrey A. Kistanov[1,*], Matti Alatalo[1], Jukka Kömi[2], Marko Huttula[1], Wei Cao[1]

[1]Nano and Molecular Systems Research Unit, University of Oulu, 90014 Oulu, Finland

[2]Materials and Mechanical Engineering Unit, University of Oulu, 90014 Oulu, Finland

**\*Email:** andrey.kistanov@oulu.fi

**Abstract**

Density functional theory based computational study has been conducted in order to investigate the effect of substitution of Cr and Co components by Si on the structure, mechanical, electronic, and magnetic properties of the high entropy alloy CrCoNiFe. It is found that the presence of a moderate concentration of Si substitutes (up to 12.5 %) does not significantly reduce the structural and mechanical stability of CrCoNiFe while it may modify its electronic and magnetic properties. Based on that, Si is proposed as a cheap and functional material for partial substitution of Cr or Co in CrCoNiFe.

**Introduction**

High-entropy alloys (HEAs) have been an active area of research in the metallic materials regime since they were first proposed by Yeh et al. [1]. The broad range of HEAs consists of alloys composed of five or more elements, within the concentration range of 5–35 at.% and they are mostly characterized by unique features such as high entropy, slow diffusion and lattice distortions [2]. Instead of the conventional alloys based on an individual solvent element, HEAs opens up a whole new world owing to the compositional complexity and possible element combinations.

Previous research on HEAs were mainly focused on various combinations of metal elements such as Al, Co, Cr, Cu, Fe and Ni [3-8]. However, it is also possible to form 'derived' alloys including other elements such as Nb, Sn etc. [9-11]. The entropy modification of the conventional alloys was also done by either replacing the solvent or the solute [12]. One such alloy is Fe-Mn-Al-Si-C system, where entropy enhancement of the solute has been conducted [13]. Another alloy similar to this system is CrCoNiFe, which has shown the potential to be tuned by addition of other alloying elements [14]. One aspect which makes CrCoNiFe less attractive is the expensive Co, and hence alloys containing reduced Co compositions has an advantage. Hence the challenge is to retain the alloy properties, when reducing the Co concentration. Many elements have been explored in altering the Co content [15,16]. CoCrFeMnNi alloys were proved to be beneficial for additive manufacturing [17], while diluted Al and Mb are predicted to influence twinning which in turn increased the mechanical properties [14]. It has been shown that the betterment of mechanical properties is possible by introducing atoms which alters the ionic to atomic radius ratio and valence electron count. The addition of Al has been shown to influence the phase changes in CrCoNiFe alloys [18], while Cu introduces high entropy effects, to benefit its application in ceramics [19]. Earlier Raabe et al [13] have investigated Fe-Mn-Al-Si-C high entropy steels, and could successfully stabilize a single homogeneous FCC phase with high tensile strength and elongations. Interestingly this high entropy steel possessed enhanced mechanical properties compared to conventional steels. Si has also been investigated in the context of light weight high entropy alloys such as $Al_{20}Be_{20}Fe_{10}Si_{15}Ti_{35}$ [20]. Further Si addition has been shown to increase the yield strength and ductility of the high entropy alloy $CoFeNiSi_x$ [21], where the change in Si concentration has shown to influence the phase structure as well. The addition of Si also significantly affected the saturation magnetization of these alloys. Though there have been some studies focusing on the positive aspects of adding Si to the quaternary alloy systems focusing on various aspects, a comprehensive analysis including structural, magnetic, electronic and mechanical properties is needed to understand further material modelling of high entropy alloys. In the present study we introduce Si to the CrCoNiFe system and study the structural, magnetic, electronic and elastic properties. Our primary aim is to replace the magnetic atoms and investigate the effect of Si towards the overall change in properties of the alloy system.

## Methods

Computational simulations were performed in the framework of the spin-polarized density functional theory (DFT) based on plane-wave method and the generalized gradient approximation (GGA) parameterized by Perdew–Burke–Ernzerhof (PBE) exchange–correlation functional [22] as implemented in the Vienna Ab initio Simulation Package (VASP) [23]. The structures were fully relaxed until the atomic forces and total energy were smaller than 0.01 eV/Å and $10^{-8}$ eV, respectively. An energy cutoff of 520 eV was adopted for the expansion of electronic wavefunctions. Periodic boundary conditions were applied in all three directions. The first Brillouin zone was sampled with k-point mesh grids of 5 × 5 × 5 (for the structure optimization) and 10 × 10 × 10 (for the electronic structure calculations) using the Monkhorst–Pack scheme. The elastic constants were derived from the strain-stress relationship as implemented in the VASP code. An analysis of elastic tensors was partly conducted in an open-source application ELATE [24].

## Results and Discussion

*Structure*. Figure 1 represents the supercell of CrCoNiFe and a schematic representation of Si substitution. Si-containing structures were created by replacing Cr or Co atoms with Si in the CrCoNiFe matrix.

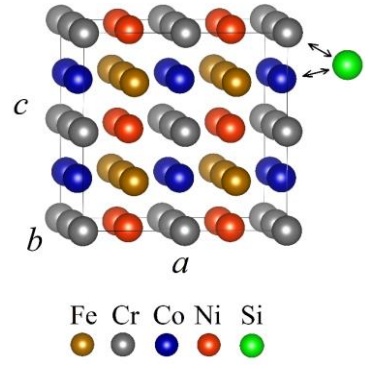

**Figure 1.** The supercell of CrCoNiFe and a schematic representation of Si substitution.

Table 1 gives the calculated equilibrium lattice constants of CrCoNiFe, $Si_xCr_{1-x}CoNiFe$, and $Si_yCrCo_{1-y}NiFe$ crystals. The calculated lattice parameters $a$, $b$, and $c$ of CrCoNiFe to be 3.492 Å for the unit cell and 6.984 Å for the 2×2×2 supercell. The obtained results are well aligned with theoretical predictions [25] and experimental observations [26, 27]. These data are also given in Table 1 for comparison.

**Table 1.** The calculated lattice parameters $a$, $b$, $c$, and formation energy $E_f$ of $Si_xCr_{1-x}CoNiFe$ and $Si_yCrCo_{1-y}NiFe$ crystals.

|  | $a$, Å | $b$, Å | $c$, Å | $E_f$, eV/atom |
|---|---|---|---|---|
| CrCoNiFe (unitcell) | 3.492 | 3.492 | 3.492 | 0.08 |
| CrCoNiFe (2×2×2 supercell) | 6.984 | 6.984 | 6.984 | 0.08 |
| CrCoNiFe (unitcell) Ref. results [26, 27] | 3.571 | 3.571 | 3.571 | 0.08 |
| $Si_xCr_{1-x}CoNiFe$ | | | | |
| x=12.5 | 7.168 | 6.992 | 6.810 | 0.01 |
| x=25 | 7.097 | 7.020 | 6.957 | 0.12 |
| x=37.5 | 7.093 | 7.025 | 6.945 | 0.26 |
| x=50 | 6.999 | 7.075 | 7.145 | 0.34 |
| $Si_yCrCo_{1-y}NiFe$ | | | | |
| y=12.5 | 7.224 | 6.991 | 6.924 | -0.09 |
| y=25 | 7.310 | 6.891 | 6.902 | -0.06 |
| y=37.5 | 7.339 | 6.888 | 6.879 | -0.03 |

The formation energy $E_f$ of CrCoNiFe and its Si-containing derivatives is defined as

$$E_f = (E_{tot} - N_{Cr}E_{Cr} - N_{Co}E_{Co} - N_{Ni}E_{Ni} - N_{Fe}E_{Fe} - N_{Si}E_{Si}) / (N_{Cr} + N_{Co} + N_{Ni} + N_{Fe} + N_{Si}) \qquad (1)$$

where $E_{tot}$ is the total energy of the unitcell, $E_{Cr}$, $E_{Co}$, $E_{Ni}$, $E_{Fe}$, and $E_{Si}$ correspond to the total energy per atom (the energy of a single atom in the bulk structure) of Cr, Co, Ni, Fe, and Si elements, respectively. The calculated $E_f$ are also collected in Table 1.

For CrCoNiFe the calculated $E_f$ of 0.08 eV/atom is similar to that reported in reference [28]. At the Si concentration of 12.5% upon the substitution of Cr by Si $E_f$ decreases almost to zero, while the increase of Si concentration to 25 % and above increases the formation energy of the $Si_xCr_{1-x}CoNiFe$ crystal. The substitution of Co by Si lowered $E_f$ of the $Si_yCrCo_{1-y}NiFe$ crystal. Particularly, at 12.5%, 25%, and 37.5% of Si $E_f$ decreases to -0.09, -0.06, and -0.03 eV/atom, respectively. Therefore, in terms the substitution of Co by Si is more energetically favourable compared to the substitution of Cr by Si. Importantly, in both considered cases, the presence of Si of a concentration of 12.5% increases the stability of the crystal. It should be noted that the formation energy of $Si_xCr_{1-x}CoNiFe$ alloys with the Si concentration higher than 12.5% is beyond 0 eV. Therefore, the introduction of Si in these cases may cause the precipitation of intermetallic phases as it has predicted for the Fe-Co-Ni-Si alloys [21].

***Elastic properties and stability***. The calculated elastic constants $C_{ij}$ of $Si_xCr_{1-x}CoNiFe$ and $Si_yCrCo_{1-y}NiFe$ crystals are collected in Table 2.

**Table 2.** The elastic constants of the orthorhombic $Si_xCr_{1-x}CoNiFe$ and $Si_yCrCo_{1-y}NiFe$ crystals, where x and y are the concentration of the element.

|  | x=0 | x=12.5 | x=25 | x=37.5 | x=50 | y=12.5 | y=25 | y=37.5 | y=50 |
|---|---|---|---|---|---|---|---|---|---|
| $C_{11}$, GPa | 386.94 | 385.27 | 339.68 | 341.93 | 315.71 | 293.48 | 340.36 | 309.57 | 285.25 |
| $C_{12}$, GPa | 196.62 | 177.56 | 179.20 | 166.71 | 168.90 | 121.49 | 156.52 | 249.76 | 244.89 |
| $C_{13}$, GPa | 196.39 | 194.78 | 175.24 | 176.92 | 148.94 | 110.44 | 148.31 | 140.81 | 153.35 |
| $C_{21}$, GPa | 196.63 | 177.56 | 179.20 | 166.71 | 168.90 | 121.49 | 156.52 | 249.76 | 244.89 |
| $C_{22}$, GPa | 370.41 | 365.29 | 303.67 | 316.62 | 272.92 | 218.60 | 287.37 | 582.64 | 172.70 |
| $C_{23}$, GPa | 175.21 | 189.98 | 158.07 | 177.25 | 149.62 | 80.29 | 171.47 | 317.91 | 224.33 |
| $C_{31}$, GPa | 196.39 | 194.78 | 175.24 | 176.92 | 148.94 | 110.44 | 148.31 | 140.81 | 153.35 |
| $C_{32}$, GPa | 175.21 | 189.98 | 158.07 | 177.25 | 149.62 | 80.29 | 171.47 | 317.91 | 224.34 |
| $C_{33}$, GPa | 323.42 | 349.57 | 281.49 | 298.15 | 238.29 | 166.38 | 266.50 | 236.22 | 314.95 |
| $C_{44}$, GPa | 201.06 | 185.16 | 152.90 | 142.32 | 120.55 | 155.45 | 162.46 | 151.29 | 139.82 |
| $C_{55}$, GPa | 152.61 | 183.58 | 168.16 | 165.50 | 141.12 | 151.66 | 135.70 | 130.06 | 98.25 |
| $C_{66}$, GPa | 191.21 | 190.51 | 164.92 | 160.14 | 136.67 | 150.31 | 169.68 | 167.65 | 146.50 |

The considered orthorhombic crystals will be mechanically stable if the following conditions are satisfied [29]:

$$C_{11}>0,$$
$$C_{11}C_{22} - C_{12}^2>0,$$
$$C_{11}C_{22}C_{33} + C_{12}C_{13}C_{23} - C_{11}C_{23}^2 - C_{22}C_{13}^2 - C_{33}C_{12}^2>0 \qquad (2)$$
$$C_{44}>0,$$
$$C_{55}>0,$$
$$C_{66}>0,$$

The calculated elastic constants suggest the orthorhombic $Si_xCr_{1-x}CoNiFe$ structure is mechanically stable at all the considered concentrations (x = 12.5%, 25%, 37.5%, and 50%) of Si impurities which replace Cr. In the case of Co substitution by Si, the orthorhombic $Si_yCrCo_{1-y}NiFe$ structure is stable at the Si concentration of y = 12.5%, 25%, and 37.5%, while at y = 50% the crystal is mechanically unstable.

The Voigt [30] and the Reuss approximation [31] are used for the calculation of a polycrystalline bulk modulus $K$ and shear modulus $G$. The Voigt bulk moduli $K_V$ and shear moduli $G_V$ of an orthorhombic crystal are given by

$$K_V = \frac{1}{9}(C_{11} + C_{22} + C_{33} + 2C_{12} + 2C_{13} + 2C_{23}), \quad (3)$$

$$G_V = \frac{1}{15}(C_{11} + C_{22} + C_{33} + 3C_{44} + 3C_{55} + 3C_{66} - C_{12} - C_{13} - C_{23}), \quad (4)$$

The Reuss bulk moduli $K_R$ and shear moduli $G_R$ moduli are given by

$$K_R = \frac{1}{(S_{11}+S_{22}+S_{33})+2(S_{12}+S_{13}+S_{23})}, \quad (5)$$

$$G_R = \frac{15}{4(S_{11}+S_{22}+S_{33})-4(S_{12}+S_{13}+S_{23})+3(S_{44}+S_{55}+S_{66})}, \quad (6)$$

where $S_{ij}$ are the elastic compliance constants and are given by $C_{ij}^{-1}$.

According to the Hill [32] approach, the effective bulk moduli $K$ and shear moduli $G$ which show the resistivity of a material to a volume change and to a shape change, respectively, are calculated as an arithmetic average of $K_V$ and $K_R$ and $G_V$ and $G_R$, respectively:

$$K = \frac{K_V + K_R}{2}, \quad (7)$$

$$G = \frac{G_V + G_R}{2}, \quad (8)$$

The Young's modulus $E$ and Poisson's ratio $v$ can be given by

$$E = \frac{9KG}{3K+G}, \quad (9)$$

$$v = \frac{3K-2G}{2(3K+G)}. \quad (10)$$

The calculated bulk modulus $K$, shear modulus $G$, Young's modulus $E$, and Poisson's ratio $v$ of orthorhombic $Si_xCr_{1-x}CoNiFe$ and $Si_yCrCo_{1-y}NiFe$ are given in Table 3.

The ductility and brittleness of materials are characterized by the Pugh's ratio [33] which is the ratio between the shear modulus and the bulk modulus ($G/K$). The critical value of the $G/K$ ratio, below which the material becomes brittle, is 0.5.

According to Table 3 for the case when Cr substituted by Si, the $G/K$ ratio slightly increases from 0.54 to 0.56 with an increase of the concentration of Si dopants from 0% to 12.5%. Further increase of the concentration of Si dopants from 12.5% to 50 %. leads to a decrease of the $G/K$ ratio from 0.56 to 0.49. However, the $G/K$ ratio exudes the value at which the material becomes brittle only at the S concentration of 50% and above. In case of Co substitution by Si, the $G/K$ ratio increases to 0.75 at the Si concentration of 12.5%, and decreases to 0.54 and 0.43 at the Si concentration of 25% and 37.5%, respectively.

According to "Frantsevich rule", materials with $v \leq 0.33$ are brittle and those with $v \geq 0.33$ are ductile [34]. Based on this criterion, it can be concluded that CrCoNiFe with $v = 0.27$ is brittle, that result is well agreed with experimental data [35] In case of Cr to Si substitution, the brittleness of $Si_xCr_{1-x}CoNiFe$ crystals slightly decreases but still exists as $v \leq 0.33$. Similarly, for the $Si_xCrCo_{1-x}NiFe$, even though an increase of the Si/Co ratio leads to an increase of $v$, the crystal remains brittle.

**Table 3.** The Voigt, Reuss, and Hill bulk moduli $K_V$, $K_R$, and $K$ and shear moduli $G_V$, $G_R$, and $G$, Young's modulus $E$, Poisson's ratio $v$, and Pugh's ratio of $Si_xCr_{1-x}CoNiFe$ and $Si_yCrCo_{1-y}NiFe$ crystals.

| | $K_V$, GPa | $K_R$, GPa | $K$, GPa | $G_V$, GPa | $G_R$, GPa | $G$, GPa | $E$, GPa | $v$ | $G/K$ |
|---|---|---|---|---|---|---|---|---|---|
| CrCoNiFe | 246.36 | 243.81 | 245.09 | 143.15 | 123.62 | 133.38 | 338.7 | 0.27 | 0.54 |
| Si$_x$Cr$_{1-x}$CoNiFe | | | | | | | | | |
| x=12.5 | 247.20 | 246.97 | 247.08 | 147.70 | 129.32 | 138.51 | 350.11 | 0.26 | 0.56 |
| x=25 | 216.65 | 214 | 215.33 | 124.68 | 104.44 | 114.56 | 291.91 | 0.27 | 0.53 |
| x=37.5 | 222.05 | 221.61 | 221.83 | 122.65 | 105.61 | 114.13 | 292.26 | 0.28 | 0.51 |
| x=50 | 195.76 | 191.58 | 193.67 | 103.63 | 87.891 | 95.76 | 246.64 | 0.29 | 0.49 |
| Si$_y$CrCo$_{1-y}$NiFe | | | | | | | | | |
| y=12.5 | 144.77 | 131.58 | 138.17 | 115.90 | 91.75 | 103.83 | 249.09 | 0.20 | 0.75 |
| y=25 | 205.20 | 203.90 | 204.55 | 121.43 | 99.947 | 110.69 | 281.32 | 0.27 | 0.54 |
| y=37.5 | 282.82 | 155.27 | 219.04 | 117.80 | 69.25 | 93.52 | 245.61 | 0.31 | 0.43 |

Cauchy pressure is another important parameter when considering whether the material has metallic or angular bonds to predict if it is ductile or brittle [35]. For the considered orthorhombic systems, the Cauchy pressure is defined as $C_{23} - C_{44}$ for the (100) plane, $C_{13} - C_{55}$ for the (010) plane, and $C_{12} - C_{66}$ for the (001) plane. The calculated values of the Cauchy pressure for the considered crystals are presented in Table 4. According to obtained single crystal elastic constants, for pure CrCoNiFe, the calculated value is -25.85 GPa for the pressure ($C_{23} - C_{44}$), 43.78 GPa for the pressure ($C_{13} - C_{55}$), and 5.41 GPa for the pressure ($C_{12} - C_{66}$). The results indicate that the bonding in the (100) plane has a strong metallic character, whereas those in the (010) and (001) planes have an angular character. The presence of 12.5% of Si substitute of Cr changes the bonding nature of CrCoNiFe, in such a way that (001) plane has a strong metallic character, while in the (010) and (100) planes bindings are of an angular character. In the case when Co is replaced by Si (concentration of 12.5%) there is a strong metallic bonding in all three (100), (010), and (001) planes. The further increase of Si concentration upon Co and Cr substitution by Si weakens the bonding within crystals as a strong angular character of bonding in all planes is observed.

Table 4. The Cauchy pressures (in GPa) of Si$_x$Cr$_{1-x}$CoNiFe and Si$_y$CrCo$_{1-y}$NiFe crystals, where x and y are the concentration of element.

| | CrCoNiFe | Si$_x$Cr$_{1-x}$CoNiFe | | | | Si$_y$Co$_{1-y}$CrNiFe | | |
|---|---|---|---|---|---|---|---|---|
| | x=0 | x=12.5 | x=25 | x=37.5 | x=50 | y=12.5 | y=25 | y=37.5 |
| (100) plane | -25.85 | 4.82 | 5.17 | 34.93 | 29.07 | -75.16 | 9.01 | 166.62 |
| (010) plane | 43.78 | 11.2 | 7.08 | 11.42 | 7.82 | -41.22 | 12.61 | 10.75 |
| (001) plane | 5.41 | -12.95 | 14.28 | 6.57 | 32.23 | -28.82 | -13.16 | 82.11 |

*Elastic anisotropy*. The formation and propagation of micro-cracks in a material are directly correlated with the elastic anisotropy of the crystal. The elastic anisotropy can be measured by shear anisotropic factors $A_a$, $A_b$, and $A_c$, where *a*, *b*, and *c* correspond to different crystallographic planes {100}, {010}, and {001}. For the orthorhombic CrCoNiFe and its Si-containing derivatives, shear anisotropic factors $A_a$, $A_b$, and $A_c$ can be calculated as follows [36]:

$$A_a = \frac{4C_{44}}{(C_{11}+C_{33}-2C_{13})}, \tag{11}$$

$$A_b = \frac{4C_{55}}{(C_{22}+C_{33}-2C_{23})}, \quad (12)$$

$$A_c = \frac{4C_{66}}{(C_{11}+C_{22}-2C_{12})}. \quad (13)$$

The shear anisotropy factors $A_a$, $A_b$, and $A_c$ as a function of the Si concentration for $Si_xCr_{1-x}CoNiFe$ and $Si_yCrCo_{1-y}NiFe$ crystals are plotted in Figure 2a. The elastic anisotropy in compressibility ($A_B$) and shear ($A_G$) can be expressed as follows [37]:

$$A_B = \frac{K_V-K_R}{K_V+K_R}, \quad (14)$$

$$A_G = \frac{G_V-G_R}{G_V+G_R}. \quad (15)$$

$A_B$ and $A_G$ as a function of the Si concentration for $Si_xCr_{1-x}CoNiFe$ and $Si_yCo_{1-y}CrNiFe$ crystals are plotted in Figure 2b.

For an isotropic crystal, the shear anisotropy factors are equal to 1, the deviation of the shear anisotropy factors from 1 is the degree of elastic anisotropy. From Figure 2a it is seen that the values of $A_a$, $A_b$, and $A_c$ are all exceeded 1 for CrCoNiFe and $A_a$ = 2.53 has the largest absolute deviation from 1, which indicates that the shear anisotropy for the {100} plane is the strongest. An introduction of Si as a substitute for Cr or Co increases the deviation of $A_c$ and decreases the deviation of $A_a$. Therefore, the shear anisotropy for the {001} plane becomes predominant.

In addition, $A_B$ and $A_G$ anisotropy factors are calculated. The deviation of the values of $A_B$ and $A_G$ from 0 is a measure of the anisotropic degree of the crystal. From Figure 2b it is seen that these values of $A_B$ and $A_G$ are nonzero and positive, and the value of $A_G$ is significantly larger than that of $A_B$ for CrCoNiFe. The substitution of Cr by Si leads to a fluctuation of the value of $A_B$ around 0, while the value of $A_G$ is significantly larger than that of $A_B$. The obtained results suggest that CrCoNiFe and its Si-containing derivatives possess high elastic anisotropy and a high anisotropic degree of shear modulus. Noticeably, the bulk modulus of CrCoNiFe is isotropic and the presence of Si instead of Cr does not affect its isotropy.

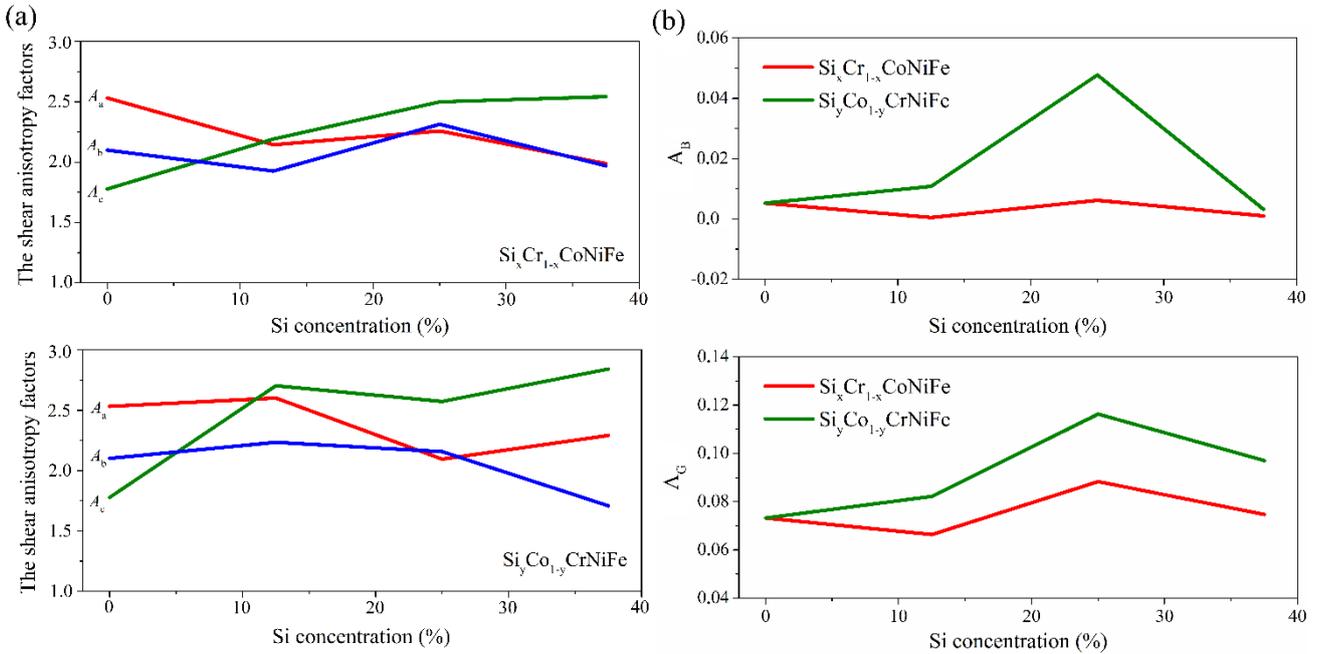

**Figure 2.** (a) The shear anisotropy factors $A_a$, $A_b$, and $A_c$ and (b) $A_B$ and $A_G$ as a function of the Si concentration for $Si_xCr_{1-x}CoNiFe$ and $Si_yCrCo_{1-y}NiFe$ crystals.

The orientation dependence of Young's modulus can also illustrate the elastic anisotropy of the crystals. For an orthorhombic crystal, the Young's modulus in any orientation is expressed as [38]

$$\tfrac{1}{E} = S_{11}\alpha^4 + (2S_{12} + S_{66})\alpha^2\beta^2 + S_{22}\beta^4 + (2S_{23} + S_{44})\beta^2\gamma^2 + S_{33}\gamma^4 + (2S_{13} + S_{55})\alpha^2\gamma^2 \quad (16)$$

where $\alpha = \sin\theta\cos\varphi$, $\beta = \sin\theta\sin\varphi$, and $\gamma = \cos\varphi$ are the direction cosines under the spherical coordinates. The orientation dependence of Young's modulus in the $xy$, $yz$, and $xz$ planes for CrCoNiFe are shown in Figure 3a–c, respectively. The plots of orientation dependence of Young's modulus in different planes at different concentrations of Si for $Si_xCr_{1-x}CoNiFe$ and $Si_yCrCo_{1-y}NiFe$ crystals are shown in Figures 4 and 5, respectively. It is seen that CrCoNiFe has elastic anisotropy because the orbit of Young's modulus in all planes is not a perfect circle. Upon substitution of Cr by Si the elastic anisotropy of the crystal is not much affected and the Young's modulus of the (100) plane has a stronger anisotropic character in comparison with other planes. The substitution of Co by Si leads to remarkable changes in Young's modulus in different planes. At the concentration of Si of 12.5% the Young's modulus of the (100) plane has a stronger anisotropic character in comparison with other planes while at the concentration of Si of 25% the Young's modulus of the (001) plane is more anisotropic. An increase of Si concentration to 50% drastically reduces the stiffness of the crystal with a Young's modulus strongly anisotropic in all planes.

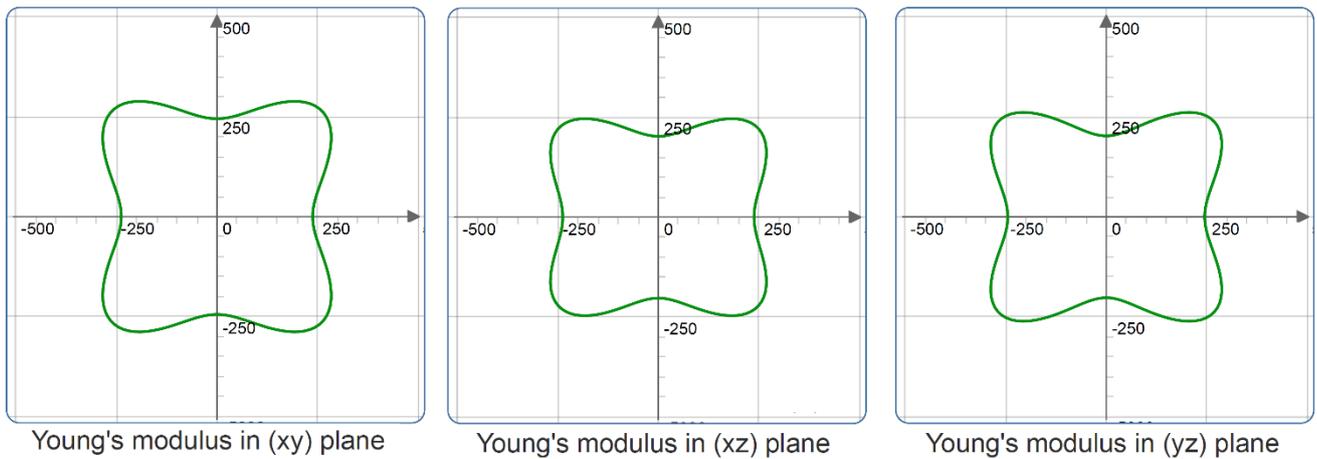

**Figure 3**. The projections of Young's modulus in the *xy*, *xz*, and *yz* planes for pure CrCoNiFe.

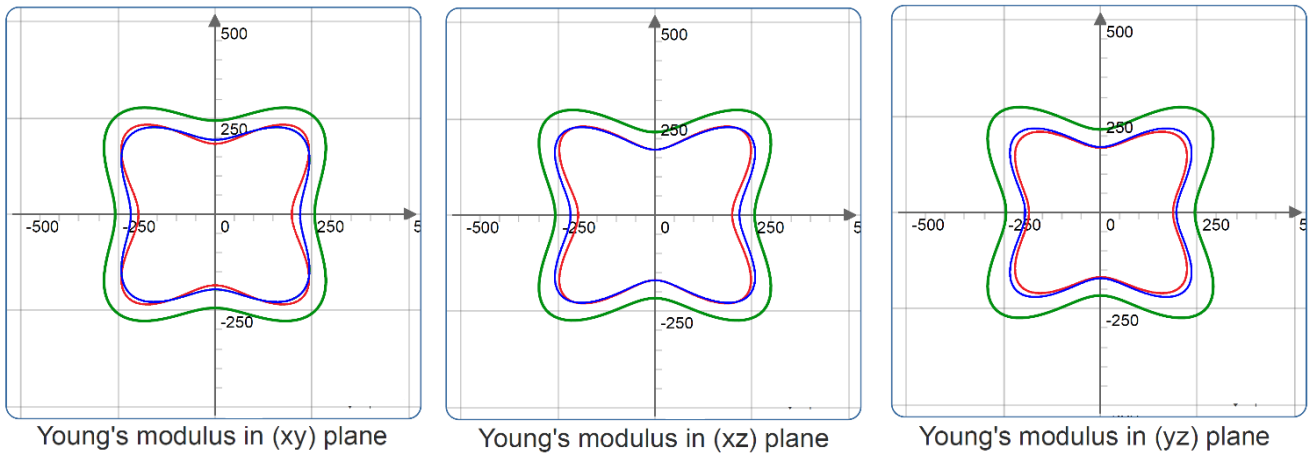

**Figure 4**. The projections of Young's modulus in the *xy*, *xz*, and *yz* planes for $Si_xCr_{1-x}CoNiFe$ at the concentration of Si of 12.5% (green line), 25% (red line), and 37.5% (blue line).

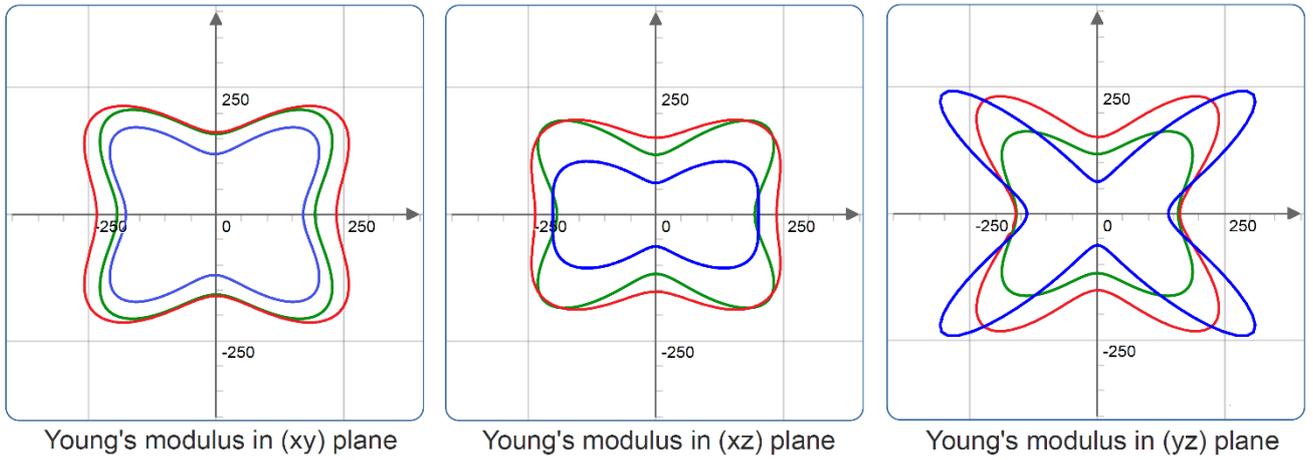

**Figure 5.** The projections of Young's modulus in the *xy*, *xz*, and *yz* planes for Si$_y$CrCo$_{1-y}$NiFe at the concentration of Si of 12.5% (green line), 25% (red line), and 37.5% (blue line).

*Electronic structure.* To investigate changes in the electronic structure of CrCoNiFe upon the introduction of Si atoms, the density of states (DOS) and partial density of states (PDOS) of each element are calculated. For the case when Cr atoms are substituted by Si atoms, Figures 6a-c show DOS of each element for the crystals CrCoNiFe, CrCoNiFe with 12.5% of Si, and CrCoNiFe with 25% of Si, respectively. The total DOSs of the three crystals exhibit metallic feature. In all cases, the bonding states are dominated by Cr-*d*, Co-*d*, Ni-*d*, Fe-*d*, and Si-*s*, *p*. A strong hybridization between Cr, Co, Ni, and Fe states implies that bonds between the atoms of these elements have a covalent character. The calculated DOSs at the Fermi level demonstrate the stability of the structures as the smaller value of DOSs at the Fermi level implies higher stability [39]. From Figure 6 it can be seen that the crystal without Si substitutes is characterized by lower density at the Fermi level compare to these with Si substitutes. With increasing the Si concentration, the density of energy states at the Fermi level is slightly increasing. These results suggest the crystal without Si substitutes suggest higher stability of that structure compared to that of with Si which is in line with our results on the mechanical stability of these structures. It should be noted that the decrease of the total DOS is caused by the decrease of intensity of Cr states. For the Si, there are both *s* and *p* states with the predomination of *s* states. According to Figure 6b, at low concentration of Si (12.5%), there is no hybridization between the main states of the crystal and Si states. Therefore, non-covalent interaction between Si and other atoms is predicted. The higher concentration of Si substitutes leads to a decrease of intensity of *s*-states of Si and a hybridization of *p*-states of Si with the states of other atoms. That may indicate covalent interaction between Si and other atoms.

For the case when Co atoms are substituted by Si atoms, Figures 7a-c show DOS of each element for the crystals CrCoNiFe, CrCoNiFe with 12.5% of Si, and CrCoNiFe with 25% of Si, respectively. The total DOSs of the three crystals also possess metallic nature. Comparing with these of pure CrCoNiFe alloy, the intensity of DOSs at the Fermi level of the Si$_x$Cr$_{1-x}$CoNiFe alloy lower at the Si concentration of 25% and significantly lower at the Si concentration of 12.5%. That may be due to a decrease of intensity of Co states which is compensated by an increase of intensity of Fe states as shown in Figure 7. Such a difference in the nature of DOSs of CrCoNiFe with Co substituted by Si compared to that of CrCoNiFe with Cr substituted by Si can explain the observed higher stability (lower formation energy) of CrCoNiFe with Co substituted by Si. For the nature of Si bonding, according to Figure 7b, at low concentration of Si (12.5%), the is no hybridization between the main states of the crystal and Si states. Hence, non-covalent bonding between Si and other atoms is predicted. The higher concentration of Si substitutes leads to a decrease of intensity of *s*-states of Si and a hybridization of *p*-states of Si with the states of other atoms. That may indicate covalent interaction between Si and other atoms.

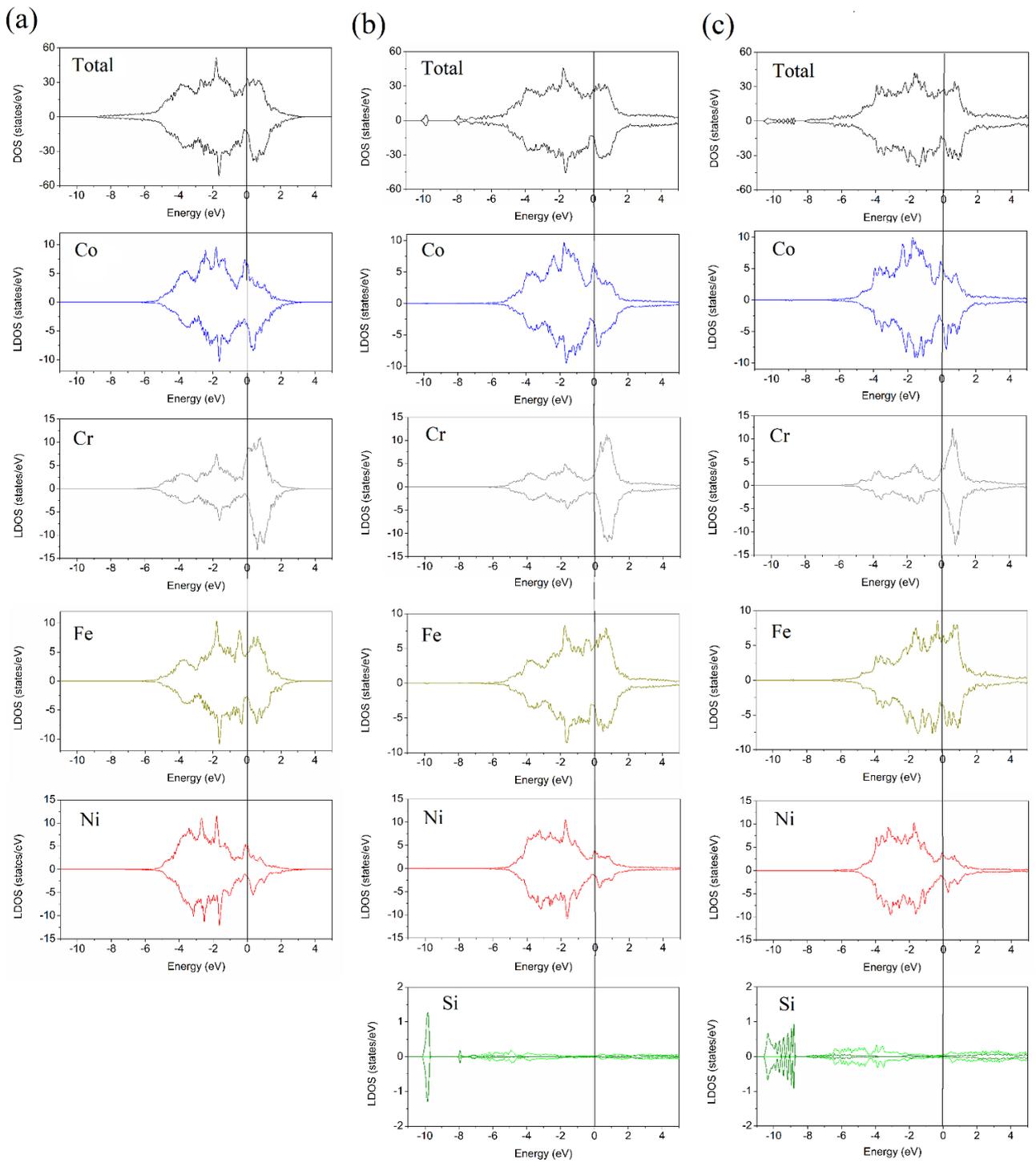

**Figure 6**. DOS and PDOS of elements of $Si_xCr_{1-x}CoNiFe$ at the concentration of Si of (a) 0%, (b)12.5%, and (c) 25%.

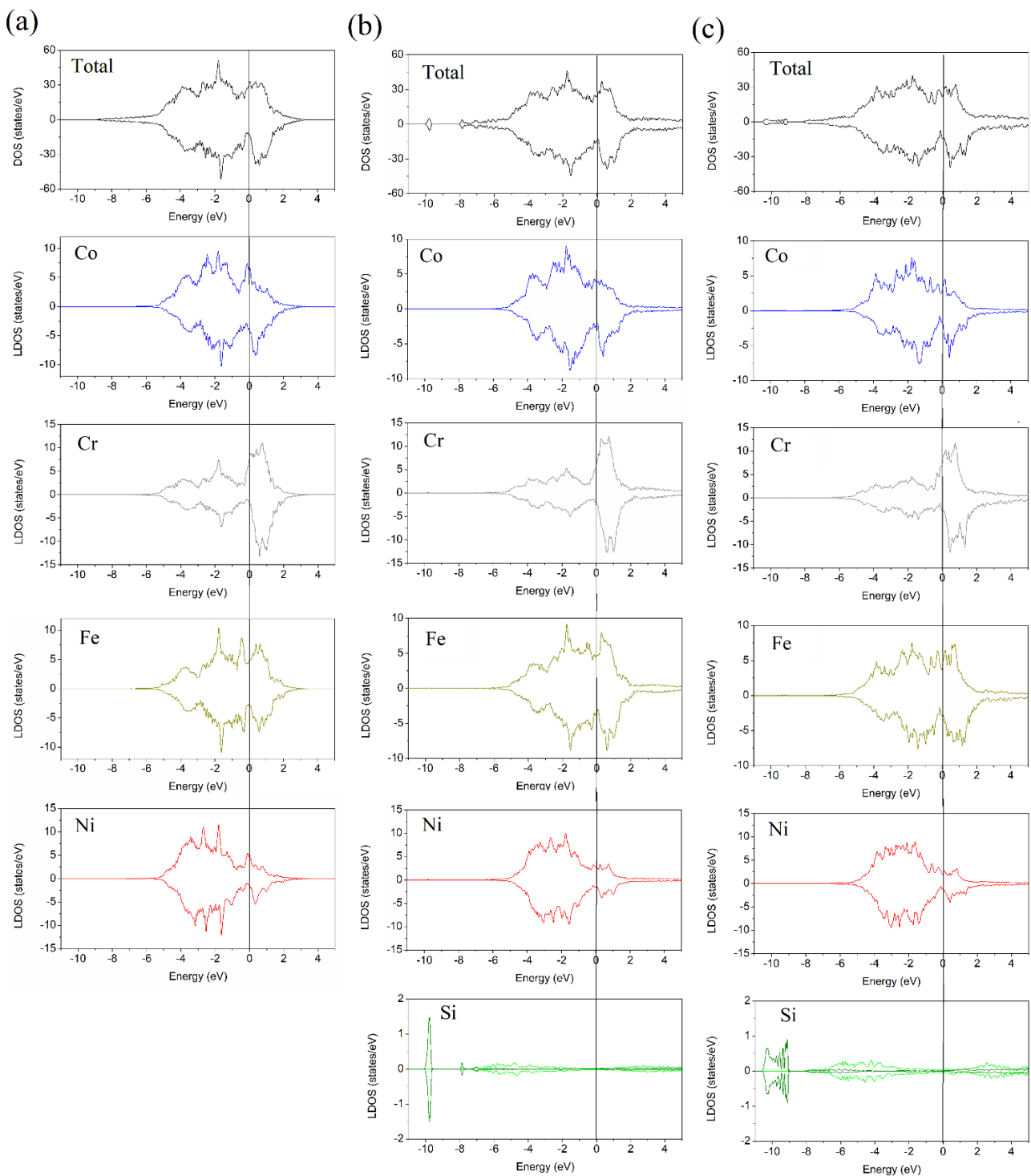

**Figure 7**. DOS and PDOS of elements of $Si_yCrCo_{1-y}NiFe$ at the concentration of Si of (a) 0%, (b)12.5%, and (c) 25%.

To get deeper understanding about the further reveal the bonding feature, characteristics, the crystal orbital overlap population (COOP) analysis is conducted for CrCoNiFe, $Si_{0.125}CrCo_{0.875}NiFe$, and $Si_{0.125}Cr_{0.875}CoNiFe$ as shown in Figure 8a-c, respectively. The negative COOP represents the occupation of the bonding state while the positive value implies antibonding [40]. It is found that there are more positive values (anti-bonding states) in CrCoNiFe than in both $Si_{0.125}CrCo_{0.875}NiFe$, and $Si_{0.125}Cr_{0.875}CoNiFe$. Therefore, $Si_{0.125}CrCo_{0.875}NiFe$, and $Si_{0.125}Cr_{0.875}CoNiFe$ possess higher stability comparing to CrCoNiFe, which is consistent with the results on the formation energy calculation. In addition, there are less positive values (anti-bonding states) of Si-Co bonds (a peak at ~-9.5 eV in Figure 8d) in $Si_{0.125}CrCo_{0.875}NiFe$ than those of Si-Cr bonds (a peak at ~-9.5 eV in Figure 8e) in $Si_{0.125}Cr_{0.875}CoNiFe$ which further explain the higher stability of $Si_{0.125}CrCo_{0.875}NiFe$ comparing to $Si_{0.125}Cr_{0.875}CoNiFe$.

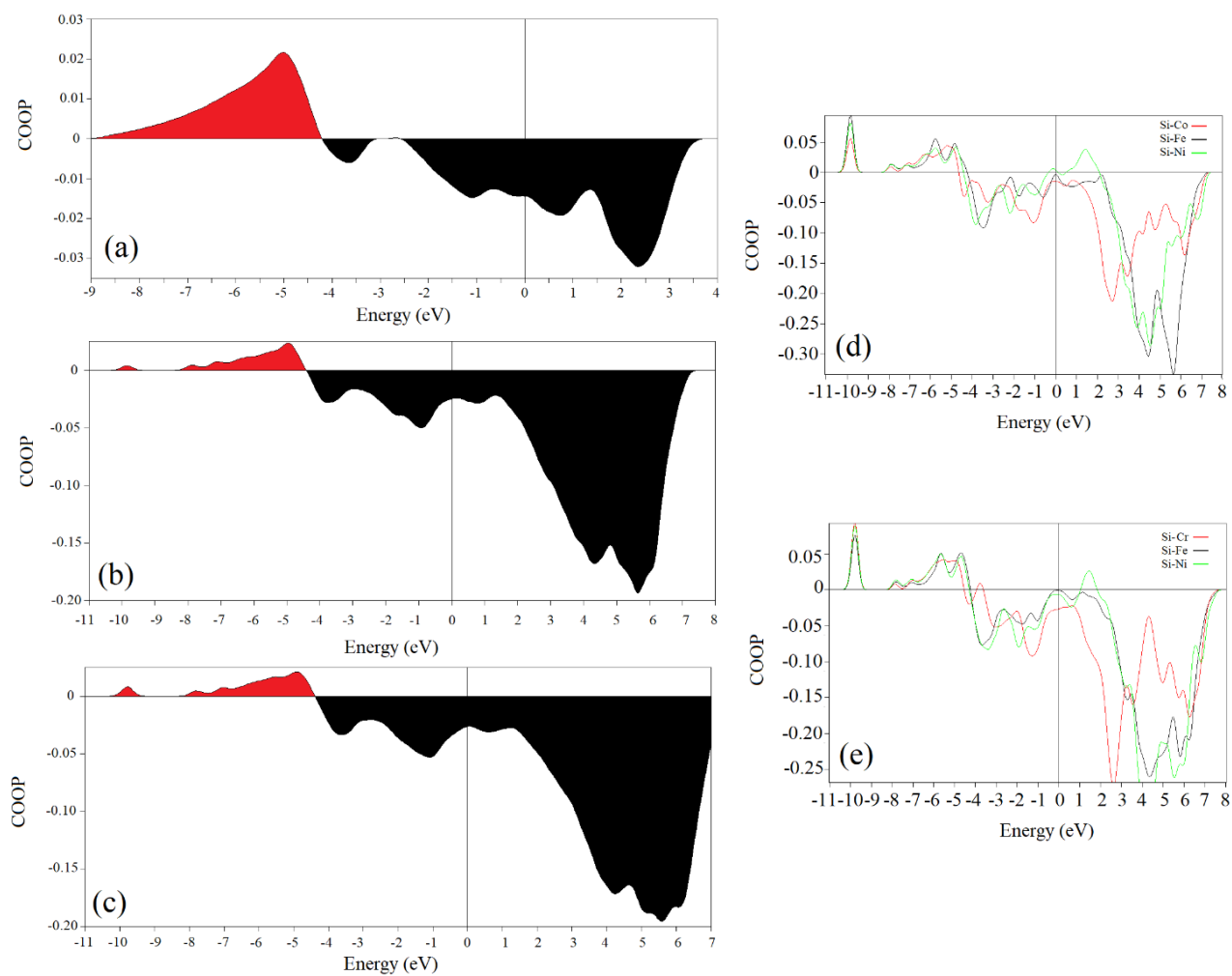

**Figure 8**. The average COOP of (a) CrCoNiFe, (b) $Si_{0.125}CrCo_{0.875}NiFe$, and (c) $Si_{0.125}Cr_{0.875}CoNiFe$. The COOP for Si in (d) $Si_{0.125}CrCo_{0.875}NiFe$ and (e) $Si_{0.125}Cr_{0.875}CoNiFe$. The Fermi level is set at zero and marked by vertical black line.

To investigate the charge transfer within the CrCoNiFe crystal upon Si-doping, the Bader analysis [41] is conducted. Table 5 collects the charges of pure elemental components of the crystal together with the charges of each element within the crystal. In pure CrCoNiFe there is a strong charge transfer (~0.96 $e$ per atom) from the Cr atoms mainly to the Ni and Co atoms. The substitution of Cr by Si decreases the loss of the charge by the Cr atoms (~0.62-0.64 $e$ per atom) while leads to the charge transfer from the Si atoms to the remaining atoms. An increase of the Si concentration lowers the charge redistribution within the crystal. The substitution of Co by Si also decreases the loss of the charge by the Cr atoms (~0.60 $e$ per atom) and the redistribution of the charge within the crystal. In addition, at the Co position the Si lose rawer electrons compared to that at the position of Cr. That also explains the significant difference in the nature of DOSs of CrCoNiFe with Co substituted by Si compared to that of CrCoNiFe with Cr substituted by Si as shown in Figures 6 and 7.

**Table 5.** The charge distribution of pure and Si-doped CrCoNiFe.

| Element (charge) | Co (9 e) | Cr (12 e) | Fe (14 e) | Ni (16 e) | Si (4 e) |
|---|---|---|---|---|---|
| | Charge of each element in the structure, e | | | | |
| | Co | Cr | Fe | Ni | Si |
| CrCoNiFe (2×2×2 supercell) | 9.34 | 11.04 | 14.13 | 16.49 | - |
| $Si_xCr_{1-x}CoNiFe$ | | | | | |
| x=12.5 | 9.14 | 11.38 | 14.10 | 16.37 | 3.74 |
| x=25 | 9.05-9.15 | 11.34 | 14.05 | 16.39 | 3.77 |
| $Si_yCrCo_{1-y}NiFe$ | | | | | |
| y=12.5 | 9.12 | 11.40 | 14.07 | 16.40 | 3.94 |
| y=25 | 9.10 | 11.40 | 14.08 | 16.43 | 3.96 |

*Magnetic properties.* To understand the preference of ferromagnetism (FM) over anti-ferromagnetism (AFM) of CrCoNiFe with Si addition the magnetic coupling energy, dE is calculated. The data is presented in Table 6.

**Table 6.** Magnetization energy (meV) of Si-doped CrCoNiFe.

| Composition | FM | AFM | dE (meV) |
|---|---|---|---|
| $Si_{0.25}Cr_{0.75}CoNiFe$ | -240.457 | -240.48 | 0.625 |
| $Si_{0.125}Cr_{0.875}CoNiFe$ | -242.531 | -243.145 | 19.18 |
| $Si_{0.125}CrCo_{0.875}NiFe$ | -244.809 | -238.521 | -196.2 |
| $Si_{0.25}CrCo_{0.75}NiFe$ | -244.032 | -239.569 | -139.43 |

The total energy is calculated by orienting spins along parallel and antiparallel directions and the magnetization energy/cell is calculated as,

$$dE = (E_{FM} - E_{AFM})/N \qquad (17)$$

where $E_{FM}$ is the total energy of the cell with the magnetic moments of all atoms initialized with positive values to obtain ferromagnetic configuration, while an antiferromagnetic configuration is simulated by initializing negative magnetic moments on some of the Cr and Si atoms. N is the total energy of atoms in the cell. From this definition, a negative value of dE indicates that FM is preferred and vice versa. It is seen that the increase of the concentration of Si dopants is leading to the stabilization of FM orientation. This indicates Si can effectively replace Co, by stabilizing FM orientation. This finding aligns with previous studies wherein change in Si concentration is related to the change in the saturation magnetization of the system [21]. It is also worthwhile to note that Si is the smallest atom, among all atoms, considered in the present study, with an atomic radius of 1.11 Å. Hence, once Si is substituted for other atoms, the volume of the cell decreases. This reduction in volume will be resulted in changes in the electron sharing and also affects the magnetic anisotropy of the system, which will result in reorientation of magnetic moments. This reorientation of magnetic moments by Si has also been observed in other materials such as Fe3Ge [42].

## Conclusions

In this work the effect of Co and Cr substitution by Si in a CrCoNiFe high-entropy alloy on its structural stability, mechanical, electronic and magnetic properties is studied. It has been revealed that the structural stability of Si-doped CrCoNiFe is higher in case of the substitution of Co by Si compared to the case of the substitution of Cr by Si. Mechanical stability of Si-doped CrCoNiFe is predicted for the concentration of Si up to 37.5% if Co is replaced and up to 50% if Cr is replaced. From the calculated Pugh's ratio, Cauchy pressure, and Poisson's ratio, it is shown that the brittleness of CrCoNiFe slightly decreases with the substitution of Cr or Co with Si. It is also found that the Young's modulus of CrCoNiFe slightly increases in the case of substitution of Cr by Si at the concentration of Si of 12.5% while the higher concentrations of Si lead to the decrease of Young's modulus of CrCoNiFe. The replacement of Co by Si leads to a drastic decrease of the Young's modulus of CrCoNiFe. According to electronic structure calculations, at a low Si concentration (12.5%) a non-covalent interaction between Si and other atoms in Si-doped CrCoNiFe is observed while the higher concentration of Si substitutes leads to a covalent interaction between Si and other atoms in Si-doped CrCoNiFe. This is also confirmed by the distribution of charges at each atom obtained through the Bader analysis. In addition, Si can effectively replace Co by stabilizing the ferromagnetic moment of CrCoNiFe. Therefore, it is proposed that Si can be a feasible alloying element for the modification of magnetic and electronic properties of CrCoNiFe. On the other hand, Si doping of CrCoNiFe does not significantly affect its mechanical properties which suggests Si is a cheap and abundant dopant for CrCoNiFe-based alloys.

## Conflicts of interest

There are no conflicts to declare.

## Acknowledgments

The authors acknowledge CSC – IT Center for Science, Finland for computational resources and the financial support provided by the Academy of Finland (grant No. 311934).